\documentclass[12pt]{article} 
 
\usepackage[]{amsmath,amssymb}
\usepackage[]{graphicx}
\usepackage[]{latexsym}
\usepackage{geometry}
\usepackage{color}
\usepackage{amscd}
\usepackage[all,cmtip]{xy}
\usepackage{mathrsfs}
\usepackage[margin=10pt,font=small,labelfont=bf]{caption}
\usepackage{color}
\definecolor{darkblue}{rgb}{0.1,0.1,.7}
\usepackage[dvips, colorlinks, linkcolor=darkblue, citecolor=darkblue, urlcolor=darkblue, linktocpage]{hyperref} 
\usepackage[square, comma, sort&compress,numbers]{natbib}
\newcommand{\reef}[1]{(\ref{#1})}
\def\beq{\begin{equation}} 
\def\eeq{\end{equation}} 
\def\beqrr{\begin{array}} 
\def\eeqrr{\end{array}} 
\def\beqa{\begin{eqnarray*}} 
\def\eeqa{\end{eqnarray*}} 
 
\font\mybb=msbm10 at 12pt 
\def\bb#1{\hbox{\mybb#1}} 
\def\bZ {\bb{Z}}

\def\b1 {\bb{1}} 

\def\del {\partial} 
\def\nn{\nonumber}

\def\lesssim{\mathrel{\hbox{\rlap{\hbox{\lower4pt\hbox{$\sim$}}}\hbox{$<$}}}} 
\def\gtrsim{\mathrel{\hbox{\rlap{\hbox{\lower4pt\hbox{$\sim$}}}\hbox{$>$}}}}

\def\en{\varepsilon}
\def\eps{\epsilon}
\newcommand{\hhref}[1]{\href{http://arxiv.org/abs/#1}{arXiv:#1}}

\numberwithin{equation}{section}

\begin{document} 

\vspace*{-.6in} \thispagestyle{empty}
\begin{flushright}
LPTENS--11/44
\end{flushright}
\vspace{.2in} {\Large
\begin{center}
{\bf Conformal Bootstrap in Three Dimensions?}
\end{center}}
\vspace{.2in}
\begin{center}
Slava Rychkov
\\
\vspace{.3in} 
\it Laboratoire de Physique Th\'{e}orique, \'{E}cole Normale Sup\'{e}rieure,\\
and Facult\'{e} de Physique, Universit\'{e} Pierre et Marie Curie (Paris VI),
France
\end{center}

\vspace{.3in}

\begin{abstract}
We discuss an idea of how 3D critical exponents can be determined
by Conformal Field Theory techniques. 
\end{abstract}
\vskip 1cm \hspace{0.8cm} November 2011

\newpage

\tableofcontents

\section{Introduction}
There are many reasons to study Conformal Field Theory (CFT): 
\begin{itemize}
\item
Beyond the Standard Model physics can contain sectors which are conformal in the UV. These can be {\it ad hoc} sectors, like unparticles, or sectors having to do with the Electroweak Symmetry Breaking, like in Walking or Conformal Technicolor.

\item Understanding of quantum gravity in AdS can be achieved by asking questions about a dual CFT. Conformal symmetry also governs structure of field correlators in dS space, relevant for inflation.

\item Finally, conformal symmetry is crucial for the theory of critical phenomena, be that in classical statistical physics, or in quantum condensed matter (`quantum criticality'). This was the primordial reason for the birth of CFT. This is also the motivation behind this paper. 
\end{itemize}

Consider the paradigmatic model of criticality, the classical $O(N)$ ferromagnet in $D=3$ dimensions, described by a partition function
\beq
Z=\exp\Bigl[ -\frac 1T \sum_{\langle ij\rangle}
\sigma_i\cdot \sigma_j\Bigr]\,, 
\label{eq:ON}
\eeq
where $\sigma_i$ are unit $N$-component vectors at the vertices of a regular cubic lattice. The simplest case $N=1$, $\sigma=\pm1$, is the Ising model with the $\bZ_2$ global symmetry.

Near and slightly above the critical temperature, the spin-spin correlation function of this model is expected to be power-like at short distances with an exponential decay at large distances:
\beq
\langle \sigma(r)\sigma(0)\rangle = \begin{cases}
r^{-2\Delta_\sigma},\quad &r \ll \xi\,,\\
r^{-(D-1)/2}\exp(-r/\xi),\quad &r \gg \xi\,,\end{cases}
\label{eq:ss}
\eeq
where $\xi=\xi(T)$ is the correlation length. At $T\to T_c$ the correlation length diverges, and the spin-spin correlator exhibits scale invariance. See \cite{Cardy} for this and other elementary facts about critical phenomena.

The emergence of conformal symmetry at the critical point is more mysterious. This seems to be a generic feature of criticality but why this happens is not fully understood \cite{polchinski}. Recently there was a renewed interest in the question whether there exist interesting scale invariant but not conformally invariant systems \cite{me, maxwell, buican, grinstein}. At this time, we don't have anything to add to this discussion and will take conformal invariance of the critical point for granted. 

\section{Critical exponents\ldots}
The most important parameters characterizing the critical point are the critical exponents. In this talk we will mostly need only two of them. The first one is the spin field dimension $\Delta_\sigma$ defined in \reef{eq:ss}.\footnote{This parameter is in one-to-one correspondence with the critical exponent commonly called $\eta$.} The second one is the energy field dimension $\Delta_\en$, defined by an analogous equation
\beq
\langle \en(r)\en(0)\rangle = 
r^{-2\Delta_\en}.
\label{eq:ee}
\eeq
The energy field $\en$ is interesting because it's the lowest-dimension scalar operator of the theory which is a singlet of the global symmetry ($O(N)$ or $\bZ_2$). This operator is relevant:
\beq
\Delta_\en <D=3\,
\eeq
(see Eq.~\reef{eq:lat} below). Adding it to the CFT action as a perturbation:
\beq
S_{CFT}\to S_{CFT}+t\, a^{\Delta_\en-3} \int d^3 x\,\en(x),\qquad t\ll 1\,,
\label{eq:pert}
\eeq
we generate an RG flow. If $t\ll 1$, then the perturbation is small at the UV cutoff length scale $a$, but it eventually becomes big in the IR, breaking conformal symmetry. The length scale $\xi$ at which this happens is determined from the equation
\beq
\xi^{\Delta_\en-3}\sim t\, a^{\Delta_\en-3}\,.
\label{eq:xi}
\eeq
This length coincides with the correlation length.

In fact, Eq.~\reef{eq:pert} is the continuum space description of what happens in the Ising model when the temperature is slightly detuned away from the critical temperature, with $t=(T-T_c)/T_c$ the reduced temperature. Eq.~\reef{eq:xi} gives then the well-known equation for the critical exponent $\nu$, which measures how fast $\xi$ grows when $T\to T_c$, in terms of $\Delta_\en$.

It is also important that $\en$ is the \emph{only} relevant singlet scalar. To reach the Ising critical point, we finetune just one parameter (the temperature), hence one relevant singlet scalar. More such scalars would imply multicritical behavior.
\section{\ldots and what is known about them}

All existing techniques to determine the 3D critical exponents are based on the idea of universality. Universality means that lots of models with different UV structure, once put at the critical point will reduce in the IR to the same CFT with the same critical exponents. The only thing which should matter is the global symmetry of the underlying Lagrangian ($\bZ_2$, $O(N)$, etc.).

One can take then any real-world system whose microscopic structure has the needed symmetry (magnets etc) and extract critical exponents experimentally, see e.g.~\cite{zj} for a summary of existing measurements.

In the same spirit, one can consider the lattice model \reef{eq:ON} and do Monte-Carlo simulations. These ``measurements" are computer time-consuming, but one has more control over interesting observables. One can also modify the lattice action to improve the numerical performance by reducing lattice artifacts etc. Using these techniques, Ref.~\cite{lattice} reports the following 3D Ising dimensions: 
\beq
\Delta_{\sigma}=0.5183(4),\qquad\Delta_{\en}=1.412(1)\qquad\text{(3D Ising)}.
\label{eq:lat}
\eeq

However, by far the most famous technique to compute the $O(N)$ 3D critical exponents has been the $\eps$-expansion \cite{4-eps}.
In this case the UV theory is the theory of $N$ free scalars $\phi^i$ perturbed by the operator $S=(\phi^2)^2$, relevant if $D<4$:
\beq
\mathscr{L}=\frac 12 (\del_\mu \phi^i)^2+\lambda\, a^{\Delta_{S}-4} S\,.
\label{eq:WF0}
\eeq
The IR theory in $D=3$ should be the $O(N)$ critical point, but to get there directly is difficult since the IR fixed point lies far away from the UV one in the theory space. Alternatively, one could say that the 3D $O(N)$ critical point does not allow a weakly coupled description in terms of the UV degrees of freedom.

The idea of Wilson and Fisher was that in $4-\eps$ dimensions for $\eps\ll1$ the $O(N)$ critical point becomes weakly coupled.
In this case the perturbing operator $S$ is only \emph{weakly relevant}:
\beq
\Delta_{S}=(4-\eps)-\eps\,. 
\eeq
The lowest order $\beta$-function takes the form
\beq
\beta_\lambda=-\eps \lambda +(N+8)\frac{\lambda^2}{2\pi^2}\,,
\eeq
and an IR fixed point lies at 
\beq
\frac{\lambda_*}{2\pi^2} \approx \frac{\eps}{N+8}\ll 1\,.
\eeq
So, in $4-\eps$ dimensions all quantities of interest (like the IR operator dimensions) can be computed perturbatively, expanding in $\eps$. For example, the first nontrivial terms for $\Delta_\phi$ and $\Delta_{\phi^2}$ are given by
\begin{align}
\Delta_{\phi}  &  =(  1-{\epsilon}/2)+\frac{N+2}{4(N+8)^{2}}\epsilon^{2}+\ldots\,,\nn\\
\Delta_{\phi^2}  &  =\left(  2-{\epsilon}\right)
+\frac{N+2}{N+8}\epsilon+\ldots\,,
\label{eq:WF}
\end{align}
where the terms in parentheses denote the UV dimensions. 
\begin{figure}[htbp]
\begin{center}
\includegraphics[scale=0.3]{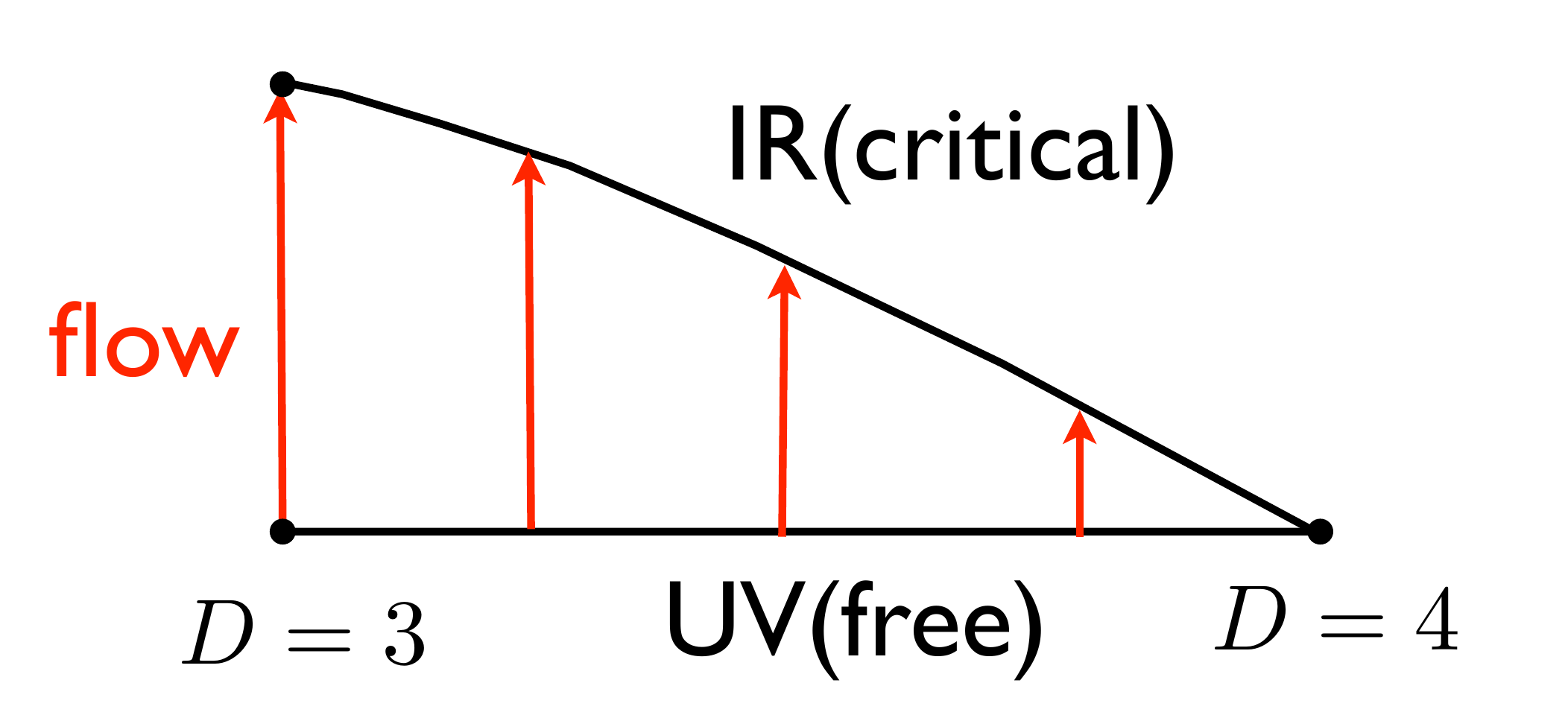}
\caption{The accepted phase diagram of the $O(N)$ fixed points for $3\le D \le 4$.}
\label{eq:phase}
\end{center}
\end{figure}

One then makes the natural assumption that this line of fixed points as a function of $\eps$ continuously connects to the 3D $O(N)$ fixed  point in the limit $\eps\to1$; see Fig.~\ref{eq:phase}. In particular, the field $\phi$ should connect to $\sigma$ and $\phi^2$ to $\en$. Setting $\eps=1$ in the series \reef{eq:WF} then allows to calculate 3D critical exponents. 

The series \reef{eq:WF} have been extended to terms of order $\eps^5$ \cite{zj}, and the 3D exponents evaluated this way are in a very good numerical agreement with experiment and Monte-Carlo studies. Notice however that as always with perturbation theory, these series are only asymptotic. For $\eps=1$, the divergent nature of the series starts to show already after the first couple of terms, so they need to be resummed (via a Borel transform), otherwise higher order terms do not improve the accuracy. The resummed results have a residual theoretical ambiguity associated with free parameters entering the resummation procedure. To quote~\cite{zj}: 

{\small ``Needless to say, with free parameters and short initial series it becomes possible to find occasionally some transformed series whose apparent convergence is deceptively good."}

\section{Why one would like to do better}

The situation described in the previous section is not fully satisfactory for the following reason. Presumably, the $O(N)$ critical points 
are very special isolated points in the space of theories. It would be nice to be able to pick out these ``diamonds" by focussing on them directly, without having to flow from free theory. After all, this is what happens in the two dimensional case, where there 
is a neat classification of critical points based on conformal symmetry \cite{BPZ}. The critical exponents of the 2D Ising model can then be determined exactly: 
\beq
\Delta_\sigma=1/8,\quad \Delta_\en=1\qquad \text{(2D Ising)}\,.
\label{eq:2Dexact}
\eeq
On the other hand, conformal symmetry is left unused in the RG calculations leading to the $\eps$-expansion. This is because this symmetry only emerges at the critical point; it's not present along the flow.

I would like to quote from a 2003 interview with A.M.~Polyakov \cite{int}, where the desire for a better 3D theory has been stated with clarity:\bigskip

{\small ``Let me tell you what I think of the renormalization group. I think there are two types of useful equations. One type is human-made, they are invented by people. The other type reflects some `pre-established harmony.' They can be discovered (uncovered) and not invented. Renormalization group is clearly a human made thing. It's clearly a smart way of calculating things but it doesn't have a breathtaking quality of, say, the Dirac equation.

The example of the second kind is operator product expansions. They form some beautiful mathematical relations and I was dreaming in the 1970s to have some classification of fixed points based on the possible operator product expansions.
The program was a little like classifying Lie algebras. In that case you start with the commutator relations which define the Lie algebra and then you classify all possible semi-simple algebras. You arrive at a stunningly beautiful theory (which was clearly discovered and not invented). I was working on that project in the 1970s and I still think it might have a chance. It was successful in two dimensions. We can classify possible fixed points in two dimensions using operator product expansions. That's what conformal field theories are about. And I think it's not excluded, that in 3 dimensions something like that is still possible. I was working for a while on this without much success in the 1970s and then I switched to other things.

I think the epsilon expansion ended the subject in the practical sense. You can calculate more or less what you want with good accuracy but aesthetically the subject is not closed yet. It's possible that there will be classification of fixed points in three dimensions, based on string theory, similar to what we have in two dimensions. But that's just dreams."}


\section{Conformal bootstrap}

We would now like to show how information about operator dimensions can be extracted by conformal field theory techniques. The idea is called ``conformal bootstrap" and is actually quite old \cite{FGG-bootstrap,pol}. However, its practical implementation has appeared only recently. In this section we will demonstrate results achieved so far in $D=4$ and $D=2$ dimensions. In the next section we will explain how the same idea applied in $D=3$ could be used to determine the $O(N)$ critical exponents. As we proceed, it should become clear why we hope to succeed where previous attempts did not have much luck.

\subsection{Old stuff}
Conformal bootstrap studies consistency conditions for four point functions. The simplest constraint comes from the four point function of the lowest dimension scalar, $\sigma$ in the Ising model case:
\beq
\langle \sigma(x_1) \sigma(x_2) \sigma(x_3) \sigma(x_4) \rangle\,.
\label{eq:4pt}
\eeq 
Conformal symmetry implies that this correlator must have the form
\beq
x_{12}^{-2\Delta_\sigma}
x_{34}^{-2\Delta_\sigma} g(u,v)\,,
\label{eq:g}
\eeq
where $x_{ij}\equiv |x_i-x_j|$ and $g(u,v)$ is a function of the conformal cross ratios 
$u=(x^2_{12}x^2_{34})/(x^2_{13}x^2_{24})$,
$v=(x_{14}^2x^2_{23})/(x^2_{13}x^2_{24})$.

A representation for this correlator and for the function $g(u,v)$ can be obtained by using the OPE. The leading terms in the $\sigma\times\sigma$ OPE are
\beq
\sigma(x) \sigma(0)=|x|^{-2\Delta_\sigma}(1+\lambda_{\sigma\sigma\en}|x|^{\Delta_\en} \en(0)+\ldots)\,,
\label{eq:sse}
\eeq
where the OPE coefficient $\lambda_{\sigma\sigma\en}$ also appears as an overall constant in the three point function
\beq
\langle\sigma(x)\,\sigma(0)\,\epsilon(y)\rangle
=\frac{\lambda_{\sigma\sigma\en}}
{|x|^{2\Delta_\sigma-\Delta_\epsilon}
|x-y|^{2\Delta_\epsilon}
|y|^{2\Delta_\epsilon}}\,,
\eeq
whose coordinate dependence is fixed by the conformal symmetry \cite{3-pt}.

Beyond the leading terms, the general conformally invariant OPE can be written as a sum over conformal primary operators $\mathscr{O}$ of the form:
\beq
\sigma(x_1)\,\sigma(x_2)
=\sum_{\mathscr{O}}\lambda_\mathscr{O}\, C(x_1-x_2,\partial_{x_2})
\mathscr{O}(x_2)\,.
\eeq
In other words, each operator $\mathscr{O}$ appears in the OPE accompanied by its derivatives. The coefficient functions $C(x_1-x_2,\partial_{x_2})$ giving these subleading contributions are fixed by conformal symmetry. There is one overall coefficient $\lambda_\mathscr{O}$ per conformal family.

In general, the $\sigma\times\sigma$ OPE will contain infinitely many primaries $\mathscr{O}=\mathscr{O}_\Delta^{(l)}$ of spins $l=0,2,4\ldots$ Only even spins will appear since we are considering the OPE of two identical scalars. The dimensions of these operators and the coefficients $\lambda_\mathscr{O}$ are at this stage free parameters. However, unitarity of the theory ($\equiv$\,reflection positivity in the Euclidean) implies that $\lambda_\mathscr{O}$ should be real, and that the dimensions of spin $l$ fields should satisfy the lower bounds \cite{unitarity4D,unitarity}:
\beq
\Delta\ge \begin{cases} D/2-1& (l=0)\,,\\
l+D-2&(l\ge 1)\,.
\end{cases}
\label{eq:unit}
\eeq

Coming back to the four point function \reef{eq:4pt}, one can apply OPE to the 12 and 34 pairs of points. One then reduces the four point functions to a sum of two point functions acted upon by the differential operators $C(x_1-x_2,\partial_{x_2})$ and $C(x_3-x_4,\partial_{x_4})$. The sum involves only diagonal terms: $\mathscr{O}=\mathscr{O}'$ since only identical fields have nonzero two point functions. It follows that the function $g(u,v)$ can be represented in the form
\beq
g(u,v)=\sum_\mathscr{O} (\lambda_\mathscr{\mathscr{O}})^2 g^{(l)}_\Delta(u,v)\,,
\eeq
where $g_\Delta^{(l)}$ are completely fixed functions of $u,v$ which depend only on the dimension and spin of the exchanged operator $\mathscr{O}$. These functions are called \emph{conformal blocks}.

We are now ready to formulate the conformal bootstrap consistency condition: the function $g(u,v)$ represented as a sum of conformal blocks should satisfy the crossing symmetry relation:
\beq
\boxed{g(u,v)=(u/v)^{\Delta_\sigma} g(v,u)}\,.
\label{eq:cross}
 \eeq
This relation follows from the fact that the four point function \reef{eq:4pt},\reef{eq:g} should be invariant under $1\leftrightarrow 3$, which corresponds to $u \leftrightarrow v$.

Alternatively, the same condition can be formulated as the associativity of the OPE. We can do the OPE either in the 12-34 channel, or in the 13-24 channel, and the result should come out the same.

\emph{A word of caution concerning terminology}: It was probably Ref.~\cite{BPZ} that first used the term ``conformal bootstrap'' in the sense just described (although the idea is 10 years older). In more ancient times the same name was sometimes used to denote a completely different thing: Feynman diagram perturbation theory with bare propagators having anomalous dimensions built in from the start. These dimensions then had to be determined from a kind of Dyson-Schwinger equation. This technique is also known as the ``skeleton expansion''. It makes sense in the large $N$ approximation, see e.g.~\cite{Petkou}, but is not useful in general. To quote classics \cite{pol}: 

{\small ``The form of the [skeleton expansion] equations depended in an essential way on the type of fundamental fields and on the form of their bare interactions, whereas the results of the theory with anomalous dimensions should not be sensitive to the choice of the initial Hamiltonian''}

\subsection{Recent stuff}
\label{sec:recent}

The bootstrap condition was first formulated in the 70s \cite{FGG-bootstrap,pol} but at the time it did not lead to solutions of 3D or 4D CFTs.\footnote{On the other hand, in 2D the bootstrap program was effectively carried out \cite{BPZ}. The simplification in 2D was that fields are grouped into bigger multiplets under the Virasoro algebra extending the finite-dimensional conformal group. In minimal models, there is only a finite number of Virasoro primaries, hence a finite number of Virasoro conformal blocks. In addition the dimensions of the fields are known. For this reason crossing symmetry constraint could be solved, determining also the OPE coefficients.} It seems that two basic difficulties were responsible for the lack of progress. First, conformal blocks were not well understood at the time; they were known only in terms of complicated integrals or power series expansions in $u,v$. The second reason is that the crossing symmetry constraint \reef{eq:cross} is objectively hard, being a functional equation for infinitely many operator dimensions and the OPE coefficients. It takes some effort to imagine how such an equation can be useful without further input. One may also fear that to get control over the theory one will have to also impose crossing for correlators of all fields rather than just $\sigma$, at which point the problem gets even more unwieldy. 
 
However, recently progress has been achieved both in finding simple expressions for conformal blocks and in extracting concrete results out of the bootstrap condition.

First, Dolan and Osborn \cite{DO1,DO2} put the $D=4$ conformal blocks in the following explicit form:
\begin{gather}
g_\Delta^{(l)}(u,v)   =\frac{(-)^{l}}{2^{l}}
\frac{z\bar{z}}{z-\bar{z}}\left[  \,k_{\Delta+l}(z)k_{\Delta-l-2}(\bar
{z})-(z\leftrightarrow\bar{z})\right]\qquad(D=4)\,,\nn\\[3pt]
k_{\beta}(z)    \equiv x^{\beta/2}{}_{2}F_{1}\left(  {\beta}/2,\beta/2,\beta;z\right)\,, \nn\\[3pt]
u=z\bar{z},\quad v=(1-z)(1-\bar{z})\,.
\label{eq:DO}%
\end{gather}

\begin{figure}[htbp]
\begin{center}
\includegraphics[scale=0.3]{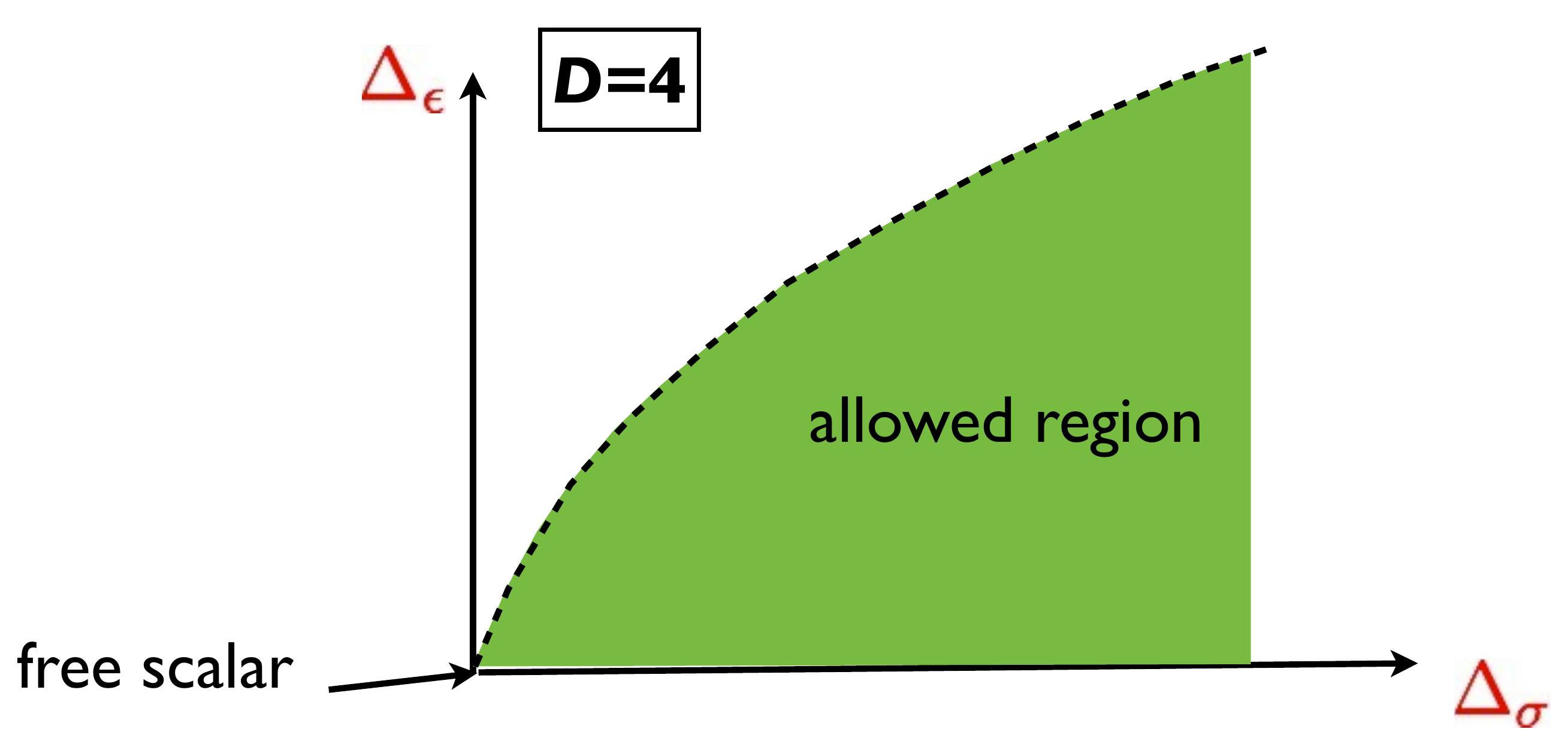}
\caption{A bound on $\Delta_\en$ as a function of $\Delta_\sigma$, shown here schematically.}
\label{fig:d4}
\end{center}
\end{figure}

Second, in \cite{us} we studied constraints imposed by the bootstrap equation on $\Delta_\en$ as a function of $\Delta_\sigma$. That study was also performed in $D=4$, motivated by some questions in electroweak phenomenology beyond the Standard Model, but this is not important for the present discussion. Rather surprisingly, we found a completely universal upper bound of the form
\beq
\Delta_\en\le f(\Delta_\sigma)\,,
\eeq
starting at the 4D free scalar theory point ($\Delta_\sigma=1$, $\Delta_\en=2$) and then growing monotonically; see Fig.~\ref{fig:d4}.
The bound was found numerically, using the explicit conformal blocks \reef{eq:DO}. Reality of $\lambda_\mathscr{O}$ (hence positivity of expansion coefficients in the conformal block representation) and the unitarity bounds \reef{eq:unit} also played an important role in the analysis. 

Since then these bounds have been strengthened by improving the numerical techniques \cite{VR}, extended to the case when a CFT has a global symmetry \cite{us2,V}, or supersymmetry \cite{DD,V}. Also, the same techniques have been used to derive universal upper bounds on the OPE coefficients and lower bounds on the CFT central charges \cite{FR,DD,us3}. Further dramatic improvement of the algorithm has been recently achieved in \cite{DDV}, where current best possible bounds of all kinds can be found.

The just mentioned bounds represent valid constraints on the landscape of possible 4D CFTs. However, it is not known at present if there are any nontrivial 4D CFTs saturating them. The situation is different in 2D, where the corresponding bound has also been computed in \cite{us}, and then with better accuracy in \cite{VR}. This was possible because the 2D conformal blocks are even simpler than in 4D \cite{DO1,DO2}:\footnote{These are of course the ``small'' $SL(2,\mathbb{C})$ conformal blocks, appropriate for comparison with 4D.  ``Big'' conformal blocks for full Virasoro algebra depend on the central charge and are much more complicated objects.}
\begin{gather}
g_\Delta^{(l)}(u,v)   \propto
k_{\Delta+l}(z)k_{\Delta-l}(\bar
{z})+(z\leftrightarrow\bar{z})\qquad (D=2)\,.
\label{eq:DO2D}%
\end{gather}

\begin{figure}[htbp]
\begin{center}
\includegraphics[scale=0.3]{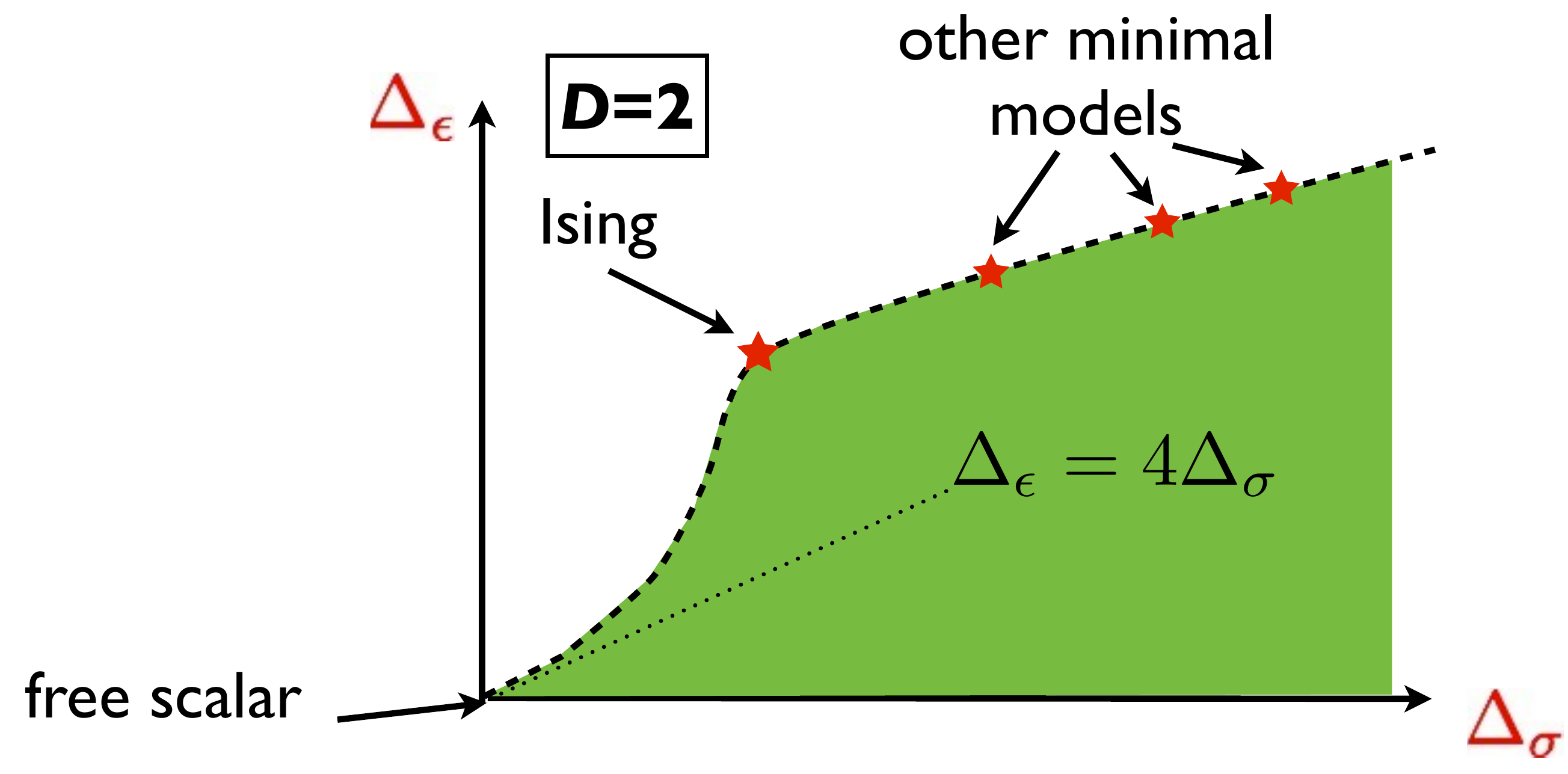}
\caption{Bound analogous to Fig.~\ref{fig:d4}, but for $D=2$.}
\label{fig:d2}
\end{center}
\end{figure}

This 2D bound is given schematically in Fig.~\ref{fig:d2} and shows interesting structure. Just as in $D=4$, the bounds begins at the free scalar point and grows monotonically. This growth starts off linear rather than as a square root, tangent to the line
\beq
\Delta_\en=4 \Delta_\sigma\,.
\label{eq:linear}
\eeq 
This fact has the following explanation \cite{us,VR}. The free scalar theory in 2D contains primary operators $V_\alpha=e^{i\alpha X}$ of dimension $\alpha^2$ (for appropriately normalized $X$) and with a leading OPE
\beq
 V_\alpha \times V_\alpha \sim V_{2\alpha}\,,\qquad V_\alpha \times V^\dagger_{\alpha}\sim \mathbf{1}\,.
 \label{eq:X}
 \eeq
 This implies an Ising-type OPE for the real parts of these operators:
\beq
\sigma \times \sigma \sim 1+\en+\ldots,\quad \sigma=\Re V_\alpha,\quad \en=\Re V_{2\alpha}\,.
\label{eq:OPEfree}
\eeq
Since the so defined fields $\sigma$ and $\en$ have dimensions related by \reef{eq:linear}, this line must lie in the region allowed by our bound. That it is actually tangent to the bound means that our result becomes best possible in the limit $\Delta_\sigma\to0$.

However, the most interesting feature of the 2D bound is that a number of solvable 2D CFTs are found to saturate it. The first of these is the 2D Ising model ($\Delta_\sigma=1/8$, $\Delta_\en=1$). The Ising point is easily identifiable since the bound has a ``knee'' here\footnote{In 4D, a similar although less pronounced ``knee'' has been noticed in a SUSY bound on dimensions of real scalars appearing in the chiral$\times$antichiral OPE \cite{DDV}. It does not correspond to any previously known theory and may be a first hint of a new SCFT.}. After the knee, the bound continues to grow linearly and passes via points corresponding to OPEs realized in higher minimal models.

\subsection{New stuff}

At this point, one may and should ask: what is the origin of the knee in the $D=2$ bound?
What is so special about $\Delta_\sigma=1/8$ that the behavior of the bound changes at this point?
We would like to show two plots which shed some light on this question. 

The basic idea is to look at the \emph{second}, after $\en$, scalar in the OPE $\sigma \times \sigma$, which we will denote $\en'$.
In the free scalar theory, the next scalar in the OPE \reef{eq:OPEfree} is marginal:
\beq
\en'=(\del X)^2,\quad \Delta_{\en'}=2\quad \text{(free 2D scalar)}\,.
\label{eq:en'free}
\eeq
On the other hand, in the 2D Ising model this scalar is strongly irrelevant:
\beq
\en'=L_{-2} \bar L_{-2}\cdot \mathbf{1},\quad \Delta_{\en'}=4\quad \text{(2D Ising)}\,.
\label{eq:en'Ising}
\eeq
These facts suggest the following exercise. For each $\Delta_\sigma$, let us fix $\Delta_\en$ to the maximal value allowed by the bound in Fig.~\ref{fig:d2}. Let us then ask what is the maximal value of $\Delta_{\en'}$ consistent with such $\Delta_\sigma$, $\Delta_\en$, and the crossing symmetry of the $\sigma$ four point function. From Eqs.~\reef{eq:en'free}, \reef{eq:en'Ising} we can suspect that such an upper bound on $\Delta_{\en'}$ should hover around 2 for $\Delta_\sigma\lesssim 1/8$, and then shoot up to about 4.

\begin{figure}[htbp]
\begin{center}
\includegraphics[scale=0.5]{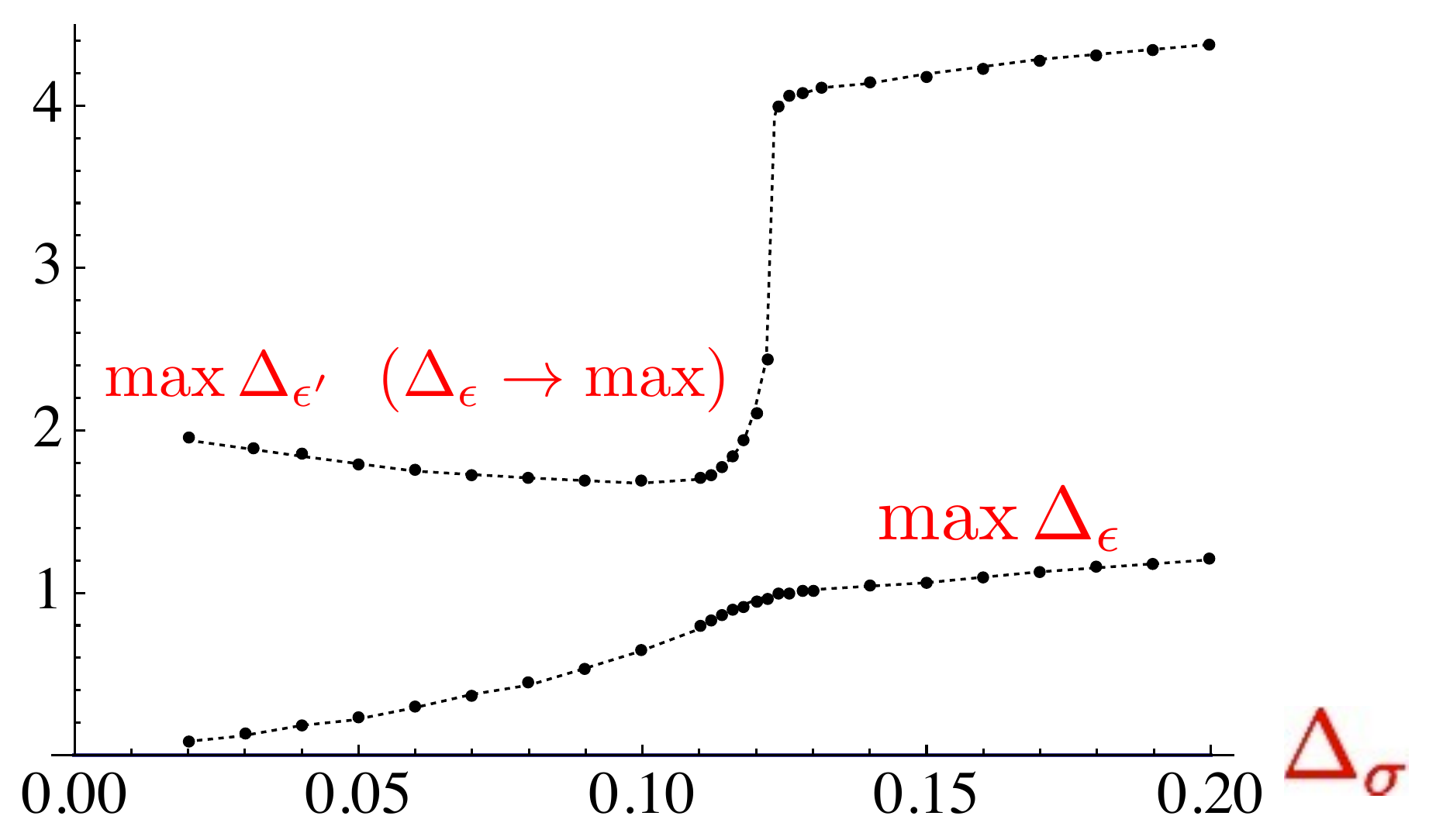}
\caption{\emph{Lower curve:} maximal possible value of $\Delta_\en$ as a function of $\Delta_\sigma$, computed from the crossing symmetry constraint by using the algorithm of \cite{VR}.${}^{\ref{note:1}}$ \emph{Upper curve:} maximal possible value of $\Delta_{\en'}$ as a function of $\Delta_\sigma$ and $\Delta_\en$ (the latter fixed to the maximal value allowed by the first bound). The dots are computed; the dashed lines are interpolated.}
\label{fig:en'}
\end{center}
\end{figure}

The bound on $\Delta_{\en'}$ can be easily computed by the algorithm used in \cite{VR} to produce the bound on $\Delta_{\en}$. Both bounds are plotted in Fig.~\ref{fig:en'}.\footnote{\label{note:1}The bounds in Fig.~\ref{fig:en'} use the linear programming algorithm in the subspace of conformal block derivatives up to order 10. The best 2D bound on $\Delta_{\en}$ presented in \cite{VR} used derivatives up to order 12 and was somewhat stronger.} The bound on $\Delta_{\en'}$ completely conforms with the above expectation: it shows a very steep, essentially step-function, growth around $\Delta_\sigma=1/8$ from $\Delta_{\en'}\approx 2$ \footnote{There is no contradiction between the bound dipping a little below 2 and the fact that $\Delta_{\en'}=2$ for free scalar. In fact $\Delta_\en$ is fixed to its maximal possible value in this exercise, which is somewhat larger than the free scalar line \reef{eq:linear}.} to $\Delta_{\en'}\gtrsim 4$.

A natural interpretation of this plot is that $\Delta_\sigma=1/8$ is the dividing line separating theories where $\en$ is the only relevant operator from theories where necessarily additional scalars with $\Delta\lesssim 2$ must be present in the $\sigma\times \sigma$ OPE.

Let us test this hypothesis further. We will study the crossing symmetry constraint demanding that there should be \emph{at most one scalar} of dimension $\le 3$ in the $\sigma\times \sigma$ OPE. In other words, we suppose that $\Delta_{\en'}\ge 3$. The cutoff value 3 here is picked somewhat arbitrarily: it is chosen to exclude the free scalar but to allow the 2D Ising model.\footnote{\label{note:gauss}Another banal theory excluded by this constraint is the ``generalized free scalar'', i.e.~a Gaussian scalar field of dimension $\Delta_\sigma$. In this case the OPE $\sigma\times\sigma$ contains operators $\en=\,:\!\sigma^2\!\!:$ and $\en'=\,:\!\!(\del^2 \sigma)\sigma\!\!:$ of dimensions $2\Delta_\sigma$ and $2\Delta_\sigma+2$, respectively.} As a consequence of a very sharp drop in the $\Delta_{\en'}$ bound of Fig.~\ref{fig:en'}, the results below will depend rather weakly on this value.

\begin{figure}[htbp]
\begin{center}
\includegraphics[scale=0.5]{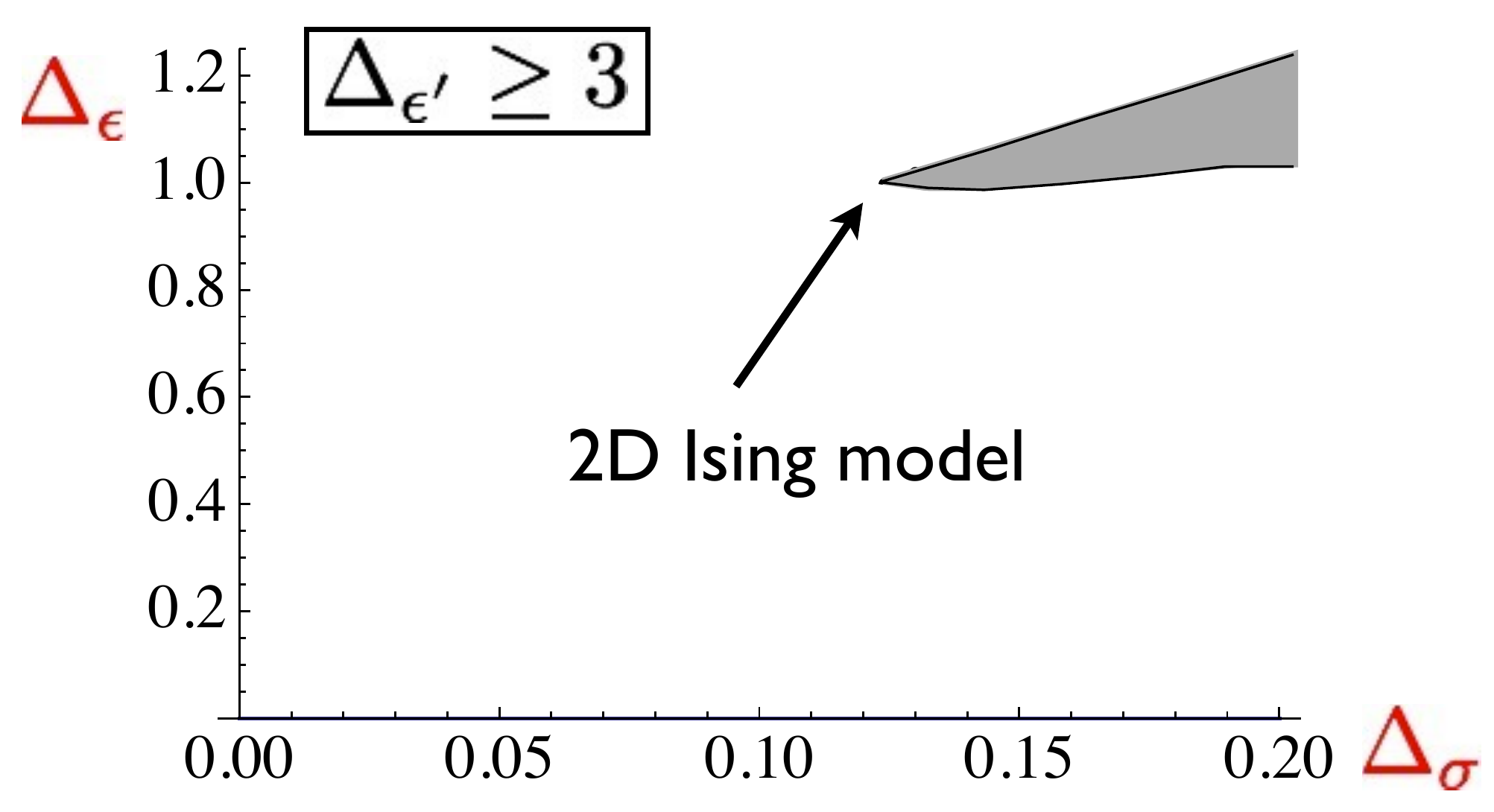}
\caption{\emph{Shaded}: the region of the $(\Delta_{\sigma},\Delta_{\en})$ plane consistent with the assumed constraint $\Delta_{\en'}\ge 3$ and the crossing symmetry in 2D CFT. Computed via the algorithm of \cite{us,VR} with derivatives up to order 10. The tip of the allowed region is at the point $\Delta_\sigma\approx0.124, \Delta_{\en}\approx0.996$.} 
\label{fig:en'>3}
\end{center}
\end{figure}

We now ask which region of the $(\Delta_{\sigma},\Delta_{\en})$ plane is consistent with the assumed constraint on $\Delta_{\en'}$ and the crossing symmetry. Once again, an answer to this question takes only a few minutes of your laptop's time to compute via the algorithm of \cite{us,VR}; it is plotted in Fig.~\ref{fig:en'>3}. 

This plot is interesting in several aspects. In marked difference with Fig.~\ref{fig:d2}, most of the allowed region goes away as a result of the $\Delta_{\en'}$ constraint. What is left is a curvy triangular region localized entirely at $\Delta_\sigma\gtrsim 1/8$. 
To be precise, the tip of the triangle is found located at the point  $\Delta_\sigma\approx0.124$, $\Delta_{\en}\approx0.996$, within 1\% from the exact 2D Ising model values \reef{eq:2Dexact}.

For $\Delta_\sigma>1/8$, the upper edge of the allowed region traces the corresponding part of the bound in Fig.~\ref{fig:d2}. Moreover, in this range of $\Delta_\sigma$ we also obtain a lower bound on $\Delta_{\en}$. The existence of this bound is a consequence of the assumed gap between the dimension of $\Delta_\en$ and $\Delta_{\en'}$.\footnote{Somewhat similarly, Ref.~\cite{DDV} has derived lower bound on the OPE coefficients of protected operators in the chiral$\times$chiral OPE as a consequence of a gap between protected and unprotected operators. In their case the gap followed from SUSY constraints on the OPE structure.} 

\emph{To summarize}: In the previous section we learned that the knee in the bound of Fig.~\ref{fig:d2} was at the 2D Ising model point. This was intriguing but not sufficiently precise to determine the $\sigma$ and $\en$ dimensions, since the knee was 
not very sharply defined. In this section we instead found that by imposing a requirement that $\en'$ be strongly irrelevant, one can get a much sharper constraint. The 2D Ising is now at the tip of the allowed region, and the dimensions are easier to extract.

What makes this approach useful is that the results depend only weakly on the assumed lower bound on $\Delta_{\en'}$. Thus we can use a very rough estimate for $\Delta_{\en'}$, like the one coming from the first order $\eps$-expansion, and still get a good determination of $\Delta_{\sigma}$ and $\Delta_{\en}$.

\section{The road to 3D}

\emph{Proposal}: The 3D Ising model critical exponents can be determined with CFT techniques. One should just redo the plots of Fig.~\ref{fig:d2} and especially Fig.~\ref{fig:en'>3} for $D=3$. One should then look for the knee and for the tip. 

As already mentioned, the lower bound on the $\en'$ dimension appropriate to use in the 3D analogue of Fig.~\ref{fig:en'>3} can be inferred from the $\eps$-expansion:
\beq
\Delta_{\en'}=(4-\eps)+\eps+O(\eps^2)\approx 4 \,.\footnote{Review \cite{vicari} gives an estimate $\Delta_{\en'}\approx 3.84(4)$ from a variety of theoretical techniques.}
\label{eq:deltaIR}
\eeq
Notice that the operator $\en'$ in this case is the $S=\phi^4$ of Eq.~\reef{eq:WF0}, which is relevant in the UV but becomes irrelevant in the IR. The previous equation refers to its IR dimension.

On the other hand in the 3D Gaussian scalar theory of the type mentioned in footnote~\ref{note:gauss}, the operator $\en'=\,:\!\!(\del^2 \sigma)\sigma\!\!:$ will have an appreciably smaller dimension $2\Delta_\sigma+2\approx 3$. Thus we should be able to differentiate between this boring theory and the 3D Ising by imposing a constraint $\Delta_{\en'}\gtrsim 3.5$. Hopefully, we will get an allowed region in the $(\Delta_{\sigma},\Delta_{\en})$ plane which will not be very sensitive to this constraint, just as it happened in $D=2$. The 3D Ising should sit at the tip of this region.

This project will need some time to carry out, because no compact representations are available for the 3D conformal blocks. Such simplified expressions as Eqs.~\reef{eq:DO}, \reef{eq:DO2D} seem to exist only in even dimensions. In $D=3$, no particular simplification was found so far, and we have to use the following rather more complicated formulas valid for generic $D$ \cite{DO1}:\footnote{See also \cite{DO3} for further work on conformal blocks in generic dimensions and in $D=3$, and \cite{CPPR} for conformal blocks of external operators with spin.}
\beq
g_\Delta^{(l)}(u,v)=u^{\frac12(\Delta-l)}f^{(l)}(b,e,\Delta;u,v),\quad b=\frac 12(\Delta-l),\ e=\frac 12(\Delta-l)\,.
\eeq
The scalar function $f^{(0)}$ is given by a double series:
\begin{align}
f^{(0)}(b,e,\Delta;u,v)&=\sum_{m,n=0}^\infty
\frac{(\Delta-b)_m (\Delta-e)_m}{m! (\Delta+1-\frac 12 D)_m}
\frac{(b)_{m+n} (e)_{m+n}}{n! (\Delta)_{2m+n}} u^m(1-v)^n\,,
\label{eq:f0}
\end{align}
where $(a)_n$ is the Pochhammer symbol. The higher-spin functions $f^{(l)}$, $l \ge 1$ are then computed from a two-term recursion relation:
\begin{align}
&f^{(l)}(b,e,\Delta;u,v)\nn\\
&=\frac{1}{2}\frac{\Delta+l-1}{D-\Delta+l-2}
\left\{
\frac{\frac D2 -e-1}{\Delta-e-1} 
\Bigl[v\, f^{(l-1)}(b+1,e+1,\Delta;u,v)-f^{(l-1)}(b,e+1,\Delta;u,v)\Bigr]\right. \nn\\
&\left. \hspace{1cm}
+\frac{\frac D2 -\Delta+e+l-1}{e+l-1} 
\Bigl[f^{(l-1)}(b,e,\Delta;u,v)-f^{(l-1)}(b+1,e,\Delta;u,v)\Bigr] \right\}\nn\\
&
-\frac{1}{4}
\frac{(\Delta+l-1)(\Delta+l-2)}{(D-\Delta+l-2)(D-\Delta+l-3)}
\frac{(\frac{D}{2}-e-1)(\frac{D}{2}-\Delta+e+l-1)}{(\Delta-e-1)(e+l-1)}
\frac{(l-1)(D+l-4)}{(\frac{D}{2}+l-2)(\frac{D}{2}+l-3)}\nn\\
 &\hspace{1cm} \times  u\, f^{(l-2)}(b+1,e+1,\Delta;u,v)\,.
 \label{eq:fl}
\end{align}
Notice that the coefficient of the second term vanishes for $l=1$, allowing to start up the recursion from just $f^{(0)}$.

To apply the algorithm of \cite{us, VR} and compute the bounds, we need to compute Taylor-expansion coefficients of the functions $f^{(l)}$ around the point $u=v=1/4$ ($\Leftrightarrow z=\bar{z}=1/2$), for any spin and for dimensions allowed by the unitarity constraints \reef{eq:unit}. To do this from \reef{eq:f0},\reef{eq:fl} looks significantly more time-consuming than the corresponding task for the 2D and 4D blocks. Nevertheless, it does not look undoable. We hope to turn to this task in the immediate future \cite{in-progress}. 

There is however a bright side to Eqs.~\reef{eq:f0},\reef{eq:fl}, following from the fact that they are valid for any $D$. The appearance of $D$ as a parameter allows us to analytically continue conformal blocks to fractional dimensions. This can be used as a starting point for a nonperturbative definition of the Wilson-Fischer fixed points and of field theory in $4-\eps$ dimensions in general. 
Notice that the crossing symmetry constraint \reef{eq:cross} is valid in any dimension. While our immediate goal is the 3D case, it would be also interesting to analyze the $D=4-\eps$ fixed points using our method. As it was noticed in \cite{us}, the naive extension of the 4D bound to $D=4-\eps$ is in contradiction with the $\eps$-expansion; it would be instructive to understand how this contradiction gets resolved.

If our approach is made to work in the simplest case of the 3D Ising model, a generalization to the 3D $O(N)$ model will also be possible. In general, in presence of a continuous global symmetry one should classify representations appearing in the OPE. There is more freedom since several representations contribute, but there are also more constraints because of the symmetry. As a result one can derive a system of equations with the same total constraining power as \reef{eq:cross}. Dimensions of each representation (and of the singlet in particular) can be bounded individually. The corresponding theory was developed in \cite{us2} and is valid in any dimension; in \cite{us2,V,DDV} the idea was shown to work in 4D. 

We argued in this talk that conformal bootstrap methods can give an alternative determination of the 3D Ising
and $O(N)$ critical exponents. While this determination will be numerical, its accuracy can be pretty high and improvable with more computer time. Importantly, we will be performing a computation which is mathematically well defined, since the OPE and \emph{a fortiori} conformal block decomposition are known to converge \cite{pol1,luscherOPE,mackOPE}. There will be no need to massage divergent series like in the $\eps$-expansion, and there will be no related resummation ambiguity. Our short term goal is to try to do better than the $\eps$-expansion for $\Delta_\sigma$ and $\Delta_\en$. The long term goal is to determine the dimensions and OPE coefficients of all the operators appearing in the $\sigma\times\sigma$ OPE.\footnote{One approach to this more ambitous task could be a nonperturbative generalization of the $4-\eps$ expansion. One would start with the free theory solution to the crossing symmetry constraint in $D=4$ and then reduce $D$ in small steps, adjusting each time the operator dimensions and OPE coefficients so that crossing symmetry (with fractional-$D$ conformal blocks) remains satisfied.} This would constitute essentially a numerical solution of the 3D Ising model at criticality.

\begin{center} 
{\bf Acknowledgements} 
\end{center}
This is a writeup of two talks given at the ``Scalars 2011'' conference in Warsaw (August 26--29, 2011) \cite{talk} and at the ``Hierarchies and Symmetries" workshop in Paris (September 18-19, 2011). I am grateful to the organizers for the hospitality.
This work is supported in part by the European Program ``Unification in the LHC Era",
contract PITN-GA-2009-237920 (UNILHC).

\end{document}